\documentclass[12pt,a4paper]{article}
\usepackage{authblk}
\usepackage[english]{babel}
\usepackage[dvipdfm]{graphicx}
\usepackage[latin1]{inputenc}
\usepackage{verbatim}
\usepackage{amsfonts}
\usepackage{amsmath}
\usepackage[T1]{fontenc}
\usepackage{color}
\usepackage{epsfig}
\usepackage{natbib}
\usepackage{url}
\usepackage[bookmarksnumbered=true, colorlinks=true, citecolor=blue]{hyperref}
\date{}

\title{A proposal for a Bohmian ontology of quantum gravity}
\author[$\dagger$$\ddagger$]{Antonio Vassallo}
\author[$\dagger$]{Michael Esfeld}
\affil[$\dagger$]{University of Lausanne, Department of Philosophy, CH-1015 Lausanne}
\affil[$\ddagger$]{University of Warsaw, Institute of Philosophy, Krakowskie Przedmie\'scie 3, 00-927 Warsaw}

\begin{document}

\maketitle
\pdfbookmark[1]{Abstract}{abstract}
\begin{abstract}
The paper shows how the Bohmian approach to quantum physics can be applied to develop a clear and coherent ontology of non-pertur\-bative quantum gravity. We suggest retaining discrete objects as the primitive ontology also when it comes to a quantum theory of space-time and therefore focus on loop quantum gravity. We conceive atoms of space, represented in terms of nodes linked by edges in a graph, as the primitive ontology of the theory and show how a non-local law in which a universal and stationary wave-function figures can provide an order of configurations of such atoms of space such that the classical space-time of general relativity is approximated. Although there is as yet no fully worked out physical theory of quantum gravity, we regard the Bohmian approach as setting up a standard that proposals for a serious ontology in this field should meet and as opening up a route for fruitful physical and mathematical investigations.
\end{abstract}

\section{The motivation for Bohmian mechanics}\label{se1}
\newcommand{\bra}[1]{\langle #1|}
\newcommand{\ket}[1]{|#1\rangle}
\newcommand{\braket}[2]{\langle #1|#2\rangle}
Basing itself on the ontology that Bohmian mechanics provides for quantum mechanics, this paper seeks to set out a proposal for a Bohmian approach to quantum gravity. We focus on loop quantum gravity, since our intention is to try out a primitive ontology of discrete beables also when it comes to a quantum theory of space-time. We start with recalling the motivation for Bohmian mechanics and its central features (this section), then sketch out how a Bohmian theory of loop quantum gravity can look like (section \ref{se2}) and finally put this theory into the larger framework of a Bohmian approach to physics in general (section \ref{se3}). Although there is as yet no fully worked out physical theory of quantum gravity, we consider the Bohmian approach as setting up a standard that any proposal for a serious ontology in this field should meet. That standard consists notably in (a) implementing a distinction between what are supposed to be elements of physical reality and what is their mathematical representation, so that the ontological commitments of the theory become clear, and (b) establishing a link between the ontological commitments of the theory and observable phenomena - in this case, a classical space-time.\\
Bohmian mechanics (BM) is a primitive ontology approach to quantum mechanics (QM) (see \citealp{203,gol}, on the notion of a primitive ontology): an ontology of matter distributed in three-dimensional space is admitted as the referent of the textbook formalism of QM, and a law for the temporal development of the distribution of matter is formulated. The motivation for doing so is to obtain an ontology that can account for the existence of measurement outcomes - and, in general, the existence of the macroscopic objects with which we are familiar before doing science.\\
BM puts forward an ontology of point-like particles that are localized in three-dimensional space. The quantum mechanical wave-function does not contain the information about the position of the particles, that is, the information about the actual particle configuration. The job of the wave-function and its temporal development according to the Schr\"odinger equation is to fix the velocity of the particles given their position. The law of BM hence is the Schr\"odinger equation for the temporal development of the wave-function and an equation, known as the guiding equation, that applies the wave-function to particle positions as initial conditions in order to obtain a velocity field for the particles:
\begin{equation}\label{1}
i\hbar\frac{\partial\Psi_{t}}{\partial{t}}=\hat{H}\Psi_{t},
\end{equation}
\begin{equation}\label{2}
\frac{dQ}{dt}=v^{\Psi_{t}}(Q).
\end{equation}
In (\ref{2}), $Q$ denotes the spatial configuration of $N$ particles in three-dimensional space and $\Psi_{t}$ the wave-function of that configuration at time $t$. More precisely, $Q$ stands for the configuration of \emph{all} the particles at time $t$, and $\Psi_{t}$ is the \emph{universal} wave-function. BM thus has a straightforward application to the universe as a whole.\\
This theory goes back to a suggestion by Louis \citet{deb}. It was later worked out by David \citet{187} and cast in an elegant manner by John Bell (see in particular \citealp{195}, ch. 17). In the following, we rely on its contemporary version as set out in \citet{323}. If one assumes that the initial particle configuration of the actual world is a typical configuration in a precise mathematical sense, one can derive the quantum mechanical probability calculus for measurement outcomes (Born's rule) from BM (see \citealp{323}, ch. 2).\\
The ontology of BM thus is the same as the ontology of classical mechanics: particles moving on definite trajectories in three-dimensional space. By contrast to classical mechanics, however, BM is a first order theory: given the position of the particles, the law fixes their velocity. The quantum mechanical features of the world are taken into account by applying to this ontology of classical objects a \emph{non-local} law: the guiding equation (\ref{2}) makes the velocity of any particle depending on, strictly speaking, the positions of \emph{all} the other particles. This is the reply that BM gives to the question of what the ontological significance of the entanglement of the wave-function in configuration space is. But there are no superpositions of values of properties in the world. The only property that the particles have is position. Since they always have a definite position, they also have always a definite value of the temporal derivative of position, that is, velocity. Given that any measurement outcome consists in something having a certain position at a certain time (such as, e.g., a pointer pointing either upwards or downwards), BM reproduces on this basis - plus the above mentioned typicality assumption about the initial particle configuration - the predictions of textbook QM for measurement outcomes. In a nutshell, BM grounds the predictions of textbook QM in an ontology of classical particles and a non-classical law.\\
Since the motivation for BM is to solve the notorious measurement problem by recognizing measurement outcomes - and, in general, localized macroscopic objects - without changing the formalism of textbook QM (the Schr\"o\-dinger equation and the Born rule), the ontological commitment of the theory is in the first place the one to particles being localized in three-dimensional space. It contradicts that motivation to conceive BM as being committed in the first place to the existence of a universal wave-function in a high-dimensional space \citep{1} or to regard it as a theory that is committed to the existence of the wave-function and that poses particles in addition to the wave-function, thus provoking the objection that the postulate of the latter is redundant (see \citealp{bro}, for that objection, and \citealp{231}, for a Bohmian reply to it).\\
The wave-function cannot be a physical entity existing in three-dimensional space in addition to the particles, pushing them to move on certain trajectories. If it is a field, it is a field on configuration space by contrast to a field in physical space. However, a causal connection between the mathematical space on which the wave-function is defined (i.e. configuration space) and physical space, with an entity belonging to the former space influencing the motion of entities existing in the latter space, would be mysterious. As mentioned above, the role of the wave-function is to fix the velocity of the particles given their position. That is to say, its function is that of a \emph{law} that yields a temporal development of something given initial conditions, the initial conditions consisting here in a configuration of particles in three-dimensional space (see \citealp{323}, ch. 12).\\
Hence, whether and in what sense the wave-function has an ontological significance in BM depends on which metaphysical stance one adopts with respect to laws. Thus, for instance, on Humeanism applied to BM, the Bohmian law (\ref{2}) with the universal wave-function figuring in it merely is the \emph{description} of the distribution of the particle positions in the whole of space-time (the Humean mosaic) that achieves the best balance between simplicity and empirical content (see \citealp{mill}, as well as \citealp{231}, section 5, and \citealp{230}, section 3). On dispositionalism applied to BM, the universal wave-function refers to or represents a holistic and dispositional property of the configuration of all the particles in the universe at any time $t$ that fixes the temporal development of the configuration, manifesting itself in the velocity of each particle at $t$ (see \citealp{202}, pp. 77-80, and \citealp{230}, sections 4-5). On primitivism about laws, over and above an initial configuration of particles, there is a fact in each possible world that a certain law holds in this world, with a certain universal wave-function figuring in that law (see notably \citealp{130} for primitivism about laws).\\
None of these stances is committed to admitting the universal wave-function in configuration space as belonging to the ontology of BM. Even on primitivism, what determines the trajectories of the particles is not the law \emph{qua} abstract entity (if abstract entities exist at all), but the law \emph{qua} instantiated in a world. However, on dispositionalism and on primitivism, by contrast to Humeanism, the universal wave-function has an ontological significance: it refers to or represents something that exists in the universe (a dispositional property, a fact) over and above the particle positions.\\
BM thus sets up a standard for a serious ontology of physics by providing, in the case of QM, a primitive ontology of matter distributed in three-dimensional space and a law for the temporal development of the primitive ontology. That standard concerns not only QM, but carries on to quantum field theory (QFT) and quantum gravity (QG). There is nothing in these latter theories given the current state of the art that allows to solve or to dispel the notorious measurement problem of QM - that is, the problem of how QM can account for measurement outcomes and in general the classical features of the world with which we are familiar (see, e.g., \citealp{324}, as regards QFT). To mention just one example in order to illustrate why such a standard is needed, consider what one of the most distinguished researchers in non-perturbative QG says about the ontology of this theory:
\begin{quote}
A weave [...] is one of many quantum states that have a certain macroscopic property, and a very peculiar one, since it is a single element of the spin network basis. There is no reason for the physical state of space not to be in a generic state, and the \emph{generic} quantum state that has this macroscopic property is not a weave state: it is a \emph{quantum superposition} of weave states. Therefore it is reasonable to expect that at small scale, space is a quantum superposition of weave states.\\
Therefore the picture of physical space suggested by LQG [loop quantum gravity] is not truly that of a small scale lattice, or as a T-shirt. Rather, it is a quantum probabilistic cloud of lattices.\\
(\citealp{4}, p. 271)
\end{quote}
Whatever the world may be, it certainly cannot be a probabilistic cloud of lattices. In the first place, probabilities always are probabilities for something, there cannot be probabilities \emph{simpliciter} in the world; when it comes to cosmological models of QG, there obviously is no question of probabilities for observations made by an observer that stands outside the system, since the system in this case is the whole universe. Moreover, a lattice is a means of representation; the question hence is what in the world the lattices employed in loop QG represent or refer to. In brief, ignoring the mentioned standard when it comes to the ontology of QG leads to proposals of which it is difficult to see what sense they could make. Let us therefore investigate whether and how a Bohmian approach can set up a standard for serious ontology also in the domain of QG.

\section{Bohmian loop quantum gravity}\label{se2}
\begin{figure}
\begin{center}
\epsfig{file=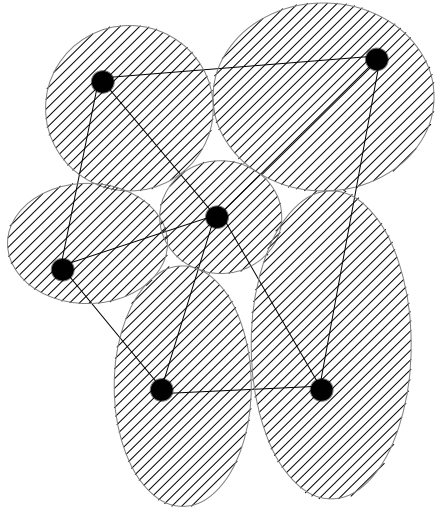,height=2.63in,width=2.21in}
\caption{An extremely simple $2$-dimensional nodes/edges-structure representing a ``bubbly'' space. The physical characterization of each bubble is encoded in the coloring of the correspondent node, while the information about ``how much'' two bubbles touch each other is conveyed by the coloring of the edges (in general, there can be more than one) linking them. Of course, in three dimensions, graphs do not have be planar.}
\label{weav1}
\end{center}
\end{figure}

\begin{figure}[h]
\begin{center}
\epsfig{file=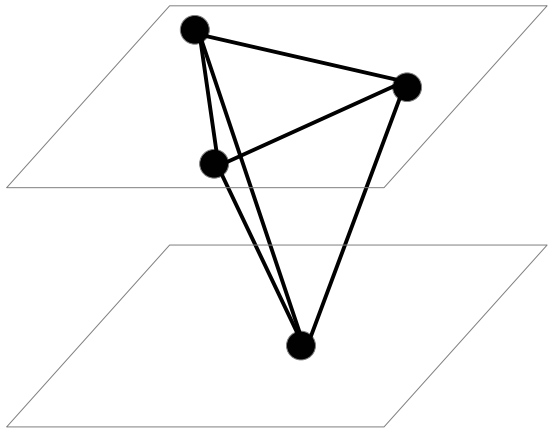,height=2.20in,width=2.80in}
\caption{Particular of a spinfoam. The dynamics expands the lower node to three contiguous nodes when moving to the subsequent upper configuration. As a result, the upper and the lower parts of the configuration are linked by three ``faces''. Such a structure is commonly called a \emph{vertex} of the spinfoam.}
\label{weav2}
\end{center}
\end{figure}

There are some sketches of Bohmian approaches to quantum gravity and quantum cosmology in the literature (e.g. \citealp{kov,squ,hol,194,192,250,193,val}). The most elaborate of these approaches from a foundational point of view is the one of \citet*{161} (reprinted in \citealp{323}, ch. 11). These authors exploit the ADM formulation of Hamiltonian general relativity theory, which consists in casting the dynamics of general relativity (GR) in terms of the evolution of a $3$-manifold (throughout the text this term will be considered synonymous with ``$3$-surface'') in coordinate time $\tau$. This is done by foliating general relativistic space-time by means of space-like Cauchy $3$-surfaces and by providing the Hamiltonian equations of motion for the canonical variables which, in this case, are the Riemannian $3$-metrics $\mathbf{h}$ defined on each $3$-surface together with the conjugate momenta depending on their extrinsic curvature. By quantizing the theory in this form - according to a procedure put forward first by \citet{139} -, the dynamics results encoded in a set of constraints over the physically allowed quantum gravitational wave-functions. Leaving aside the so-called diffeomorphism constraints, which just select all wave-functions $\Psi(\mathbf{h})$ that are invariant under smooth surface deformations ($3$-diffeomorphisms), the relevant dynamics for a canonical theory of quantum gravity is given by the Wheeler-DeWitt equation:
\begin{equation}\label{3}
\hat{H}\Psi(\mathbf{h})=0,
\end{equation}
where $\hat{H}$ is the Hamiltonian operator defined in the physical Hilbert space of the theory.\\
In order to cast this theory in a Bohmian framework, Goldstein and Teufel choose the components $h_{ab}$ of a $3$-metric as the primitive ontology, thus fixing the configuration space of their Bohmian theory of QG to be the space $Riem(\Sigma)$ of Riemannian $3$-metrics on a $3$-manifold $\Sigma$. They then propose the following guiding equation for the primitive ontology:
\begin{equation}\label{4}
\frac{dh_{ab}}{d\tau}=N(\tau)G_{abcd}Im\Bigg[\Psi(\mathbf{h})^{-1}\frac{\delta\Psi(\mathbf{h})}{\delta h_{cd}}\Bigg],
\end{equation}
where $Im[\dots]$ represents the imaginary part of the formula in parentheses, $G_{abcd}$ is the ``metric'' defined over $Riem(\Sigma)$, and $N(\tau)$ is the lapse function which, roughly speaking, encodes information on how to ``pile up'' the different $\mathbf{h}(\tau)$ to obtain a general relativistic $4$-metric.\\
There are at least four important points in Goldstein's and Teufel's proposal that we would like to highlight. The first one is that this theory is background independent in a general relativistic sense: it is not formulated on a background space-time. Instead, it treats the metrical-gravitational degrees of freedom as dynamical features. Secondly, although the Wheeler-DeWitt equation (\ref{3}) is timeless in the sense that it does not depend on any time parameter - be it physical or just mathematical -, the guiding equation (\ref{4}) generates a non-trivial dynamics for the primitive ontology. The third point is that, contrary to common expectations, in this quantum theory of gravity, there is nothing discrete about space or space-time: the primitive ontology is given by (components of) continuous $3$-metrics, and the space-time obtained from the dynamical evolution equation (\ref{4}) is continuous at all scales. Finally, the procedure of gluing of $3$-surfaces encoded in (\ref{4}) depends on the specification of a lapse function $N(\tau)$. This means that (\ref{4}) in principle selects a privileged foliation of space-time and, hence, fixes a distinguished decomposition of the $4$-geometry into an assembly of $3$-geometries. Consequently, this Bohmian theory of QG is committed to more space-time structure than admitted in GR for which, simply speaking, the only physically relevant geometrical structure is the $4$-geometry of space-time. One possibility to avoid this consequence would be to show that, in some way, the theory deals just with wave-functions that generate, through (\ref{4}), a dynamics which does not depend on $N(\tau)$; however, it is difficult to see how this idea could be implemented.\\
In contrast to the approach that Goldstein and Teufel take, we think that there is a good (Bohmian) motivation to retain the commitment to a primitive ontology of discrete objects also when it comes to quantum gravity. In putting forward an ontology of particles for QM, BM subscribes to a primitive ontology of discrete objects. In the same vein, when moving from QM to QFT, in what is known as Bell-type Bohmian QFT, the primitive ontology of particles - and thus discrete entities - is maintained so that a theory is set out in which the empirical predictions of textbook QFT are grounded in an ontology of particles (see \citealp{195}, ch. 19, and \citealp{323}, ch. 9-10; see furthermore \citealp{stru}, for an overview of the state of the art in Bohmian QFT). Hence, when passing from QFT to QG in a Bohmian spirit, there is a \emph{prima facie} good motivation to try out a primitive ontology of discrete objects also when it comes to QG. However, there obviously is no longer a question of an ontology of particles moving in a background space in QG. If the quantum regime is applied to space-time itself and if the framework of a primitive ontology of discrete objects is carried on from BM and Bell-type Bohmian QFT to QG, this means that one has to develop a primitive ontology of discrete objects for space-time itself. Such a commitment can be boosted by several arguments from physics - as, for example, the argument for the finiteness of black hole entropy (see, e.g., \citealp{352}, thesis 2) - suggesting that there is some kind of discrete structure underlying general relativistic space-time. In any case, given that the situation in QG is currently open, we take it to be worthwhile to consider a primitive ontology of discrete entities also in the domain of QG. \\
Of course, in a Bohmian context, the motivation for committing oneself to a primitive ontology of discrete objects cannot derive from operators and discrete spectra of eigenvalues of operators. The Bohmians emphasize with good reason that one cannot go from operators to ontology, but that ontology has to come first and that the deduction of empirical predictions by means of introducing operators or observables has to be done on the basis of the ontology (see notably \citealp{214}). The motivation for a primitive ontology of discrete objects thus has rather to be situated in the context of the venerable tradition of atomism in Western thought and its tremendous success in classical physics, as well as chemistry and molecular biology.\\
When trying out a primitive ontology of discrete objects for QG, it is reasonable to focus on loop quantum gravity (LQG), for this approach is committed to the view that space-time is discrete at the fundamental level. It usually subscribes to this commitment for reasons that the Bohmians reject, namely the discrete spectra of the volume and area operators defined on the Hilbert space of gravitational states. However, this fact does not rule out the possibility to develop a clear ontology for LQG in Bohmian terms. Moreover, it is of a genuine interest of its own right to explore a Bohmian approach to LQG, since LQG is the best candidate for a theory of non-perturbative QG given the current state of the art. Indeed, there are good reasons to expect that a full theoretical development of LQG will deliver a genuine general relativistic quantum theory; this expectation is strengthened by the recent results that indicate that the formalism of LQG is derivable from GR both through a straightforward canonical quantization and via a covariant quantization on a lattice (historically, one of the first works highlighting the compatibility between a canonical and a covariant dynamics for LQG is that of \citealp*{317}). Nonetheless, many details have still to be filled in - the full specification of a \emph{physical} Hilbert space of the theory, in the first place -, and a lot of technical difficulties have to be addressed (see \citealp{4}, for the standard textbook in LQG).\\
Given these open physical issues and given the fact that no work at all has been done as yet on a Bohmian approach to LQG, we have to limit ourselves in this section to sketching out how the central features of the Bohmian approach can be applied to LQG - that is, a primitive ontology and a law for the primitive ontology. In what follows, we will restrict ourselves to the case of pure gravity; that is to say, we will not take matter into consideration.\\
Since LQG is intended to be a straightforward quantum theoretical version of GR, it is a background independent theory (see \citealp{60}, for an argument why a theory of QG should be background independent, and see \citealp{311,56}, for a philosophical discussion of background independence). The - at the present stage just kinematical but, hopefully, also physical - Hilbert space of quantum gravitational states of LQG admits a countable basis formed by states called \emph{spin networks}. By specifying an embedding in a $3$-manifold, a given spin network gains a spatial representation in terms of a nodes/edges-structure, namely, a graph. Each node and edge in a graph is labelled or ``colored'' by a representation of the $SU(2)$ group, hence \emph{spin} network. The important point is that it does not matter \emph{how} a spin network is embedded in a $3$-manifold. This means that the physical information encoded in the corresponding labelled graph is not affected by arbitrary smooth deformations of the $3$-manifold. In short, we will refer to spin networks as equivalence classes of graphs under $3$-diffeomorphisms.\\
The most common interpretation of spin networks is the geometric one which exploits the fact that the labeling of nodes and edges in a graph represents contributions to eigenvalues, respectively, of volume and area operators defined on the Hilbert space of gravitational states (see \citealp{291}, and \citealp{4}, pp. 269-270, who make clear that this is not an initial assumption, but a consequence of the theory). LQG thus is committed to fundamental discrete structures underlying a classical smooth $3$-geometry (this justifies the talk of ``weave states'' often found in the literature). This commitment suggests that the complete theory has to describe the spatial geometry as being made up of discrete (perhaps Planck-sized) extensions of space that are small enough to be ``smoothed out'' when looked at from a large-scale perspective. Even though this geometrical interpretation of gravitational states in LQG is not the only possible one (see e.g. \citealp{290}, section 11.1, for some alternative accounts), it is the one that admits the most straightforward ontological reading.\\
Hence, in a Bohmian version of this theory - let us call it Bohmian loop quantum gravity or BLQG -, the objects that are introduced as the primitive ontology of the theory have to meet three requirements: (1) they cannot be local beables in the standard sense, that is, objects localized in space-time, such as unextended particles occupying a position (see \citealp{195}, ch. 7, for the notion of local beables). (2) They have to be discrete. (3) They have to approximate a smooth $3$-geometry when grouped together in a suitable way. In other words, it has to be possible to derive a smooth $3$-geometry by suitably coarse-graining a configuration of such objects.\\
Therefore, if we want to retain a commitment to Bohmian ``particles'' - i.e. discrete fundamental beables - in a background independent context, we need to switch to a view of particles as \emph{partless} objects, that is, mereological atoms. In this way, we give up a characterization of these objects in terms of a position in space while retaining their discreteness and their fundamentality. If we consider a $3$-space as a mereological whole, then the commitment to atomism implies that this whole cannot be divided in smaller and smaller regions \emph{ad libitum}. Instead, the process of division comes to an end when a mereological atom is reached. Let us therefore pose elementary extensions of space - atoms of space so to speak - as the primitive ontology of BLQG. These atoms are not \emph{localized in} space. They rather are \emph{localizations of} space. In this manner, we seek to extend the classical notion of local beables to the quantum-gravitational regime.\\
Let us now turn to configurations of these fundamental objects. In order to obtain a configuration, we have to introduce a certain fundamental relation in which atoms of space have to stand. Since we seek to recover the metrical-gravitational field from configurations of atoms of space, this relation should not be a \emph{metrical} one from the outset; it should be prior to any metrical characterization.  Let us therefore consider a set $\mathcal{X}$ of $N$ atoms of space and let us introduce a contiguity relation\footnote{Perhaps, it would be better to call it ``companionship'' relation, since the word ``contiguity'' might mistakenly suggest that such a relation bears some spatial connotation.} $C\subset\mathcal{X}\times\mathcal{X}$ such that it is:
\begin{itemize}
\item[-] \emph{Irreflexive}: $\forall x\in\mathcal{X}\neg C(x,x)$.
\item[-] \emph{Symmetric}: $\forall x,y\in\mathcal{X}(C(x,y)\equiv C(y,x))$.
\item[-] \emph{Serial}: $\forall x\in \mathcal{X}\exists y\in \mathcal{X} C(x,y)$.
\end{itemize}
From the ordered couple $(\mathcal{X},C)$ a topological space $X$ can be constructed such that its points are the elements of $\mathcal{X}$, and each subset $\{x,y\}\subset\mathcal{X}$ for which $C(x,y)$ holds is identified with the unit interval $[0,1]$. If we let two elements in $E=\{\{x,y\}\subset\mathcal{X}|C(x,y)\}$ be ``glued'' together if and only if they have (at least) one point in common, then the topology defined over $X$ turns out to be the \emph{graph topology}, which means that $X$ can be pictorially represented as a finite graph $\Gamma_{X}=(\mathcal{X},E)$ with $\mathcal{X}$ the set of nodes, and $E$ the set of edges. The coloring of the graph is then mathematically represented by a suitably defined function that assigns a label (e.g. a representation of the $SU(2)$ group) to each element of the graph.\\
We can then define in $X$ an approximation function $k_{N}$ to a ``target'' smooth $3$-manifold $\mathcal{M}$ as a funtion $k_{N}:X\longrightarrow\mathcal{M}$ depending on the number $N$ of atoms of space such that, in the ``continuum'' limit $N\longrightarrow\infty$ it becomes a homeomorphism (this is just a heuristic sketch: see \citealp{sor}, especially sections 2 and 4, for a mathematically rigorous articulation of this approximation procedure). By the same token, all the large-scale metrical quantities should be recoverable as results of the same limiting procedure on appropriately defined approximation functions (which, in general, would depend also on the coloring of a graph). Of course, implementing such an approximation is in general a highly non-trivial task (for example, for some $X$, it might be impossible to define continuum approximation functions), but it is not a desperate one. Moreover, fulfilling this task can be easily given a concrete physical meaning. Simply speaking, in fact, for a sufficiently large number of atoms of space, the small-scale picture of BLQG is that of a cluster of ``bubbles'' touching each other. Such a picture can be compactly expressed in terms of a graph where a node represents a bubble (its color representing its ``size'') and the colored edges connecting two nodes represent ``how much'' the corresponding bubbles touch each other (see figure \ref{weav1}).
\\
Not surprisingly, this procedure closely resembles the introduction of a weave state in LQG, but an important point has to be clarified in this respect. In fact, LQG conflates under the designation ``weave state'' two quite distinct concepts, namely, (a) the properly intended quantum state $\ket{X}$ (belonging to the Hilbert space) and (b) its pictorial representation $\Gamma_{X}$ (belonging to the configuration space of graphs), but it only accords full physical dignity to the former. The Bohmian reading, instead, turns the tables by regarding $\Gamma_{X}$ as standing for a \emph{concrete} configuration of \emph{concrete entities} which are the physical referents of the formalism.\\
Moreover, in standard LQG, a generic quantum gravitational state is considered to be a superposition of weave states, thus giving rise to the ontological confusion expressed in the quotation from Rovelli in the previous section. The Bohmian approach provides ontological clarity for LQG by maintaining that there are no superposed states of anything existing in nature. Superpositions enter into the calculation of probabilities for measurement outcomes, but they do not belong to the ontology of the theory. The central ontological claim of BLQG is that there is exactly one configuration of atoms of space. Consequently, BLQG does not face a principled problem in recovering the classical space-time of GR from configurations of atoms of space, since the ontology of the theory is classical from the outset, although space is discrete at the fundamental level according to this theory. By contrast, if one admits superpositions of such configurations as the ontology of LQG, then it is entirely mysterious how such superpositions could disappear in order to give rise to the classical space-time of GR. This is, of course, the way in which the notorious measurement problem of QM strikes LQG, if one does not take care to start with a primitive ontology of the theory. Furthermore, as we have seen, BLQG provides the possibility for giving a clear physical sense to the claim that a universal configuration of atoms of space ``weaves up'' a large scale smooth $3$-geometry, and, in the same manner, it provides a (yet-to-be implemented) conceptually straightforward mechanism for the appearance of physical metrical quantities at large scales resulting from an underlying non-metrical regime.\\
However, the claim that atoms of space bear no metrical properties while extended space does so may sound odd. But consider the following analogy: it is perfectly possible for a lump of matter to instantiate a certain property, say having a particular shape, while the material atoms (the particles) that compose it do not have a shape; grouping these material atoms together in a suitable manner makes it that the resulting object (the lump) has a certain shape. The same reasoning applies to atoms of space and the regions they form: grouping the atoms of space together in a suitable manner as represented by nodes and edges on a graph makes it possible for the configuration to instantiate metrical properties, while the individual atoms of space are connected only by a contiguity relation.\\
Having obtained a clear ontological characterization of the universal configuration $X$ in BLQG, we can now turn to supplying a guiding equation for its dynamics. Before doing so, let us recall that, since LQG is derived from a canonical quantization of GR, its dynamics is constrained by the Wheeler-DeWitt equation, which, in the present context, reads as follows:
\begin{equation}\label{5}
\hat{H}\Psi_{\Gamma}=0.
\end{equation}
The subscript $\Gamma$ of the (universal) wave-function means that it is defined over the configuration space of graphs. With this clarification in place, we can introduce the following - tentative - guiding equation for BLQG:
\begin{equation}\label{6}
\dot{X}\propto F\Big(\Psi_{\Gamma},\mathfrak {D}\Psi_{\Gamma}, X\Big),
\end{equation}
The $\dot{X}$ represents the evolution of a configuration of atoms of space $X$ with respect to a coordinate time $\tau$ - that is, a mathematical index that labels configurations - and $F$ is a function of $X$, of the wave-function and of $\mathfrak{D}\Psi_{\Gamma}$, that is, a suitably defined first-order derivative of such a wave-function. As a Bohmian law, (\ref{6}) should show the same mathematical form of (\ref{2}) and (\ref{4}), namely:
\begin{equation}\label{7}
\dot{X}\propto\frac{j^{\Psi_{\Gamma}}}{\rho^{\Psi_{\Gamma}}}.
\end{equation}
This means that in BLQG, as in any other Bohmian theory, the dynamical development of a configuration of atoms $X$ is related through a suitably defined coupling constant to the ratio between the probability current $j^{\Psi_{\Gamma}}=Im[\Psi_{\Gamma}^{*}\mathfrak{D}\Psi_{\Gamma}]$ and the probability density $\rho^{\Psi_{\Gamma}}=\Psi_{\Gamma}^{*}\Psi_{\Gamma}=|\Psi_{\Gamma}|^{2}$. However, at the present stage, the theoretical developments of both LQG and BLQG make it difficult to be more explicit about (\ref{6}) because we do not have at hand a clear mathematical characterization of quantum-gravitational wave functions and, most importantly, of the configuration space $S_{\Gamma}$ of the theory. For example, we do not have any clue of what topology $S_{\Gamma}$ should have: if this space turns out to be discrete, then ``topology'' might not even be the right word. Moreover, if $S_{\Gamma}$ is discrete, then (\ref{6}) and (\ref{7}) might have to be replaced by stochastic equations or, perhaps, the dynamical evolution should be considered as taking place in discrete steps, that is, $\tau$ could be a discrete parameter. This lack of implementation, then, prevents us from accomplishing the mathematical task of constructing probability densities and currents for BLQG, and hence specify (\ref{6}). Nonetheless, we can already sketch out how the dynamics of such a theory should work.\\
First of all, as in the case of QM, the dynamics of BLQG shifts the physical accent from the Hilbert space of universal gravitational states to the space of universal configurations of the elementary objects of the theory - in this case, atoms of space arranged through a contiguity relation, instead of particles occupying a position in a background space. Secondly, in this context, the universal wave-function retains, \emph{mutatis mutandis}, the twofold meaning of the quantum mechanical one. More precisely, if $\Psi_{\Gamma}$ is a solution of (\ref{5}), $|\Psi_{\Gamma}|^{2}$ provides a probability distribution over all the actualisable configurations of atoms of space at a certain dynamical stage $\tau$  - contrary to the odd standard interpretation which takes it to be the probability distribution of outcomes for an experiment performed on the entire universe ``from the outside''. Such a probability distribution does not depend on coordinate time, so it is the same for all $\tau$'s. On the other hand, $F$ singles out a unique path in the configuration space corresponding to the dynamical evolution described by (\ref{6}). This latter equation is a non-local law exactly as (\ref{2}) because all the space atoms in a configuration are entangled together by the wave-function.\\
Before clarifying what non-locality means in the present context, let us spell out in more detail what equation (\ref{6}) is supposed to do. Simply speaking, the guiding equation should encode a set of simple rules that can be illustrated in terms of ``local moves'' on graphs (expansions or contractions by means of additions or subtractions of nodes and edges). By applying and reiterating such moves on a ``starting'' graph, we generate a structure made up by a ``pile'' of graphs and, at the same time, we ``propagate'' the coloring of the starting graph throughout the structure. Such a structure does not only display labelled nodes and edges, but also ``faces'' which encode information on how the contiguity relation changes from one graph to the subsequent one (see figure \ref{weav2}).
\\
In the standard theory, such a structure is called ``spinfoam'', and the covariant dynamics of LQG is in fact a sum-over-spinfoams in the following sense. Given an initial configuration state $\ket{X_{i}}$ and a final one $\ket{X_{f}}$, the theory considers all the possible spinfoams, that is, all dynamically possible paths connecting the initial and the final configuration, assigns a probability amplitude to each path, and then calculates the total state transition amplitude $W_{if}$ as a sum over all possible paths (see e.g. \citealp{4}, section 9.1).\\
In contrast to the above mentioned dynamical descriptions, in BLQG the dynamics encoded in (\ref{6}) describes exactly one actual path in configuration space leading from the initial to the final configuration, which is fully determined once the initial configuration is plugged into equation (\ref{6}). Provided that (\ref{5}) and (\ref{6}) are really compatible, BLQG then grounds the probability amplitudes of the standard theory by taking them to express our ignorance of the final state in the configuration space given our ignorance of the initial configuration of atoms of space. In other words, BLQG adds to standard LQG a parameter - namely that there always is exactly one actual and uniquely labelled configuration of atoms of space $X$ - and provides a guiding equation for this actual configuration. In this manner, exactly one evolution path in the configuration space is singled out, and the dynamics of the theory acquires a clear ontological significance, namely to describe the actual evolution of the initial configuration of atoms of space.\\
In fact, the dynamics as given by (\ref{6}) deals with configurations as a \emph{whole} and, hence, without specifying the entire initial configuration, we cannot predict with certainty how a proper part of such configuration will evolve, because the dynamics of BLQG is non-local. As in BM, so also in BLQG, the dynamical evolution of any proper part of a configuration depends on all the other parts. To be more precise, in BM the non-locality resides in the fact that, at each instant, the velocity of a given particle assigned by (\ref{2}) depends, strictly speaking, on the positions of all the other particles, however far apart in space they may be. Of course, in BLQG, (\ref{6}) does not assign velocities to the atoms of space in a configuration and is not concerned with ``trajectories'' of atoms of space  - or, indeed, an identity of individual atoms from one configuration to the next one - but, rather, assigns to each atom a move such that a transition from one configuration to the next one is achieved. In this context, to say that the dynamics is non-local means that, for each configuration, the move assigned to a given atom depends on the way in which all the atoms are arranged, even those not contiguous with the atom considered.\\
Moreover, given that (\ref{6}) provides a dynamics for the primitive ontology, BLQG has the means to describe a physical process that involves change, namely, change in the instantiation patterns of the contiguity relation from one configuration to the subsequent one, as encoded in the change of labeling between configurations. In other words, since BLQG postulates two laws - the Wheeler-DeWitt equation (\ref{5}) and the guiding equation (\ref{6}) -, the fact that the wave-function figuring in (\ref{5}) is stationary does not create a problem for BLQG to take change into account. Hence, BLQG is in the position to restore a familiar dynamical picture in which the classical space-time of GR is approximated by a stack of universal configurations $\{X_{i}\}$ whose ordering is given by (\ref{6}). The above mentioned evolution can be conceived as a series of configurations (the $X_{i}$'s) that are ``interlaced'' by means of faces according to the guiding equation (\ref{6}) in order to form a classical space-time. The specific configurations that are involved in this dynamical process are ``selected'' by $\Psi_{\Gamma}$ according to (\ref{5}): The microscopic dynamics is linked to the macroscopic classical one by the fact that, once we apply the dynamical law to a configuration which is typical for a wave-function compatible with (\ref{5}), we generate an evolution that, at large scale, resembles the $3+1$ evolution of Hamiltonian GR. This means that each configuration approximates a smooth $3$-manifold and (\ref{6}) ``piles up'' such ``quasi'' $3$-manifolds, thus approximating a foliation of a $4$-manifold.\\
If this view proves sound, then it will be possible to recover a classical picture of a globally hyperbolic space-time which admits a (in general, non-unique) global time function (as, for example, in the Friedmann-Lema\^itre-Robertson-Walker model of GR). In such a case, there would be a clear sense in which, at a classical level, a weak temporal ordering supervenes on the fundamental ordering established by (\ref{6}): in this respect, BLQG is some sort of ``discrete cousin'' of Goldstein's and Teufel's BQG. It is important to note that, although the above dynamical picture is far from being physically implemented, it points to a conceptually clear direction that physicists can follow in seeking a concrete physical mechanism for the appearance of general relativistic space-time from the underlying quantum gravitational regime.\\
The sketch of BLQG that we have drawn in this section comes with a series of provisos, which can be summarized in the following conditional: \emph{if} a definite theory of LQG shows the features we have mentioned (in particular the ``geometrical weaves'' interpretation on which much of our reasoning was based) and \emph{if} it is possible to implement a guiding equation (\ref{6}) - be it deterministic (as we have tacitly assumed so far) or stochastic - that is compatible with the dynamics of standard LQG, then the account of configurations of atoms of space developed in this section provides a primitive ontology for this theory. Of course, it might turn out that LQG will develop in a totally different way or that, eventually, it will prove itself to be a theoretical dead end. While we have to wait for the further theoretical development of LQG, we nonetheless claim already at this stage that we have shown a way how to formulate a clear ontology of this theory, avoiding the confusions that are unfortunately widespread even in the best physical literature (cf. the quotation from Rovelli at the end of the previous section).

\section{Bohmian quantum gravity within the Bohmian approach to physics}\label{se3}
As the presentation in the preceding section makes clear, the structure of a Bohmian ontology for QG is the same as in the case of QM: one poses a primitive ontology, consisting in an initial configuration of beables, and a law for the transition from one such configuration to another one. The primitive ontology is classical in the sense that it simply exists - there are no superpositions of possible configurations of beables, just one actual configuration. What distinguishes the ontology of a quantum theory from the ontology of a classical theory is situated exclusively in the law of the development of the configuration of beables: that law is non-local in that it applies only to the configuration as a whole. It cannot be separated into factors that provide for a development of each element of the configuration taken individually (the only way to do so is by means of approximation procedures like the one that yields effective wave-functions). In a nutshell, the theory poses classical objects and a non-classical law for their development. Probabilities then enter into the theory as in classical statistical mechanics: we need probabilities only because of our - principled - ignorance of what the actual initial configuration of beables is.\\
When moving from QM via QFT to QG in a Bohmian approach, more stress is laid on the configuration of the beables and on what fixes its development than on the individual beables. In Bohmian QM, it would be wrong-headed to conceive what is represented by the wave-function as \emph{generating} the temporal development of the particles. As in Newtonian mechanics, the particles move anyway (or are at rest), and the role of what the wave-function stands for is only to fix the \emph{form} of their temporal development - in other words, to fix what the trajectories of the particles are like, but not to generate the fact that there are trajectories at all. In Bell's proposal for a Bohmian QFT, the particles (the local beables) lose ontological weight so to speak, since the particle number no longer is an invariant (see \citealp{195}, ch. 19, and \citealp{323}, ch. 10): what fixes the temporal development of the particle configuration is such that it includes stochastic events of particle creation and annihilation. Consequently, the particles (the local beables) no longer have an identity in time independently of what fixes the temporal development of the configuration. In BLQG, the individual atoms of space have even less ontological significance, a space-time is built up through the guiding equation (\ref{6}) applied to universal configurations of such atoms of space. More precisely, if we regard the universal wave-function as encoding information about all the actualisable graphs in a possible world selected by (\ref{5}) (which in the standard interpretation is taken to be a static quantum superposition of gravitational states), then (\ref{6}) establishes an actualization ordering of the universal configurations of such atoms of space. Consequently, there is nothing that connects the individual atoms of space in different configurations apart from what is supplied by what the universal and stationary wave-function stands for.\\
As explained in section \ref{se1}, the Bohmian approach consists in providing a primitive ontology for a physical theory and a law for the development of the elements of the primitive ontology. In any Bohmian quantum theory, the wave-function is not a physical entity in addition to and on a par with the elements of the primitive ontology, but falls on the side of the law: its job is to fix the development of the elements of the primitive ontology, given an initial configuration of these elements. However, this does not imply that the wave-function does not have an ontological significance. Only if one adopts a Humean attitude to laws, the ontology is exhausted by the primitive ontology, the law (and whatever figures in it) being merely an economical description of the total arrangement of the elements of the primitive ontology in a world.\\
By contrast, if one grounds laws in properties of the elements of the primitive ontology, then our proposal for BLQG results in the following ontology: the initial configuration of atoms of space instantiates a holistic and dispositional property that manifests itself in the transition from one configuration to the subsequent one. That property is represented by the universal wave-function, and it grounds the law of the development of the configurations. The fact that the universal wave-function is stationary means that the instantiation of that holistic and dispositional property is the same for any configuration of atoms of space in a world - in other words, that property does not change when moving from an initial configuration to subsequent ones.\\
In the same vein, on primitivism about laws, over and above there being an initial configuration of atoms of space, there is the fact instantiated in any world of BLQG that a certain law holds in this world, with a certain universal and stationary wave-function figuring in that law. Given the fact that a certain law is instantiated and the initial configuration of atoms of space, a development towards certain subsequent configurations of atoms of space ensues. Hence, in any Bohmian quantum theory including QG, the Bohmian framework of a primitive ontology and a law for the development of the elements of the primitive ontology allows for these elements to instantiate a property or a fact that grounds the law and that is represented by the wave-function, although it is an inaccurate description that has led to much confusion in the literature to present the Bohmian theory as being committed to the existence of the wave-function over and above the existence of the primitive ontology.\\
There is a well-known tension between any Bohmian quantum theory and relativity physics (see \citealp{198} for a precise examination of the tension between quantum non-locality and relativity physics in general). In the case of BLQG, this tension consists in the fact that, as the theory was formulated in the previous section, the transition from an initial configuration of atoms of space to subsequent ones such that the ordering of these configurations through the guiding equation (\ref{6}) approximates the space-time of GR singles out a preferred foliation of space-time by means of $3$-surfaces, as in the proposal for BQG by Goldstein and Teufel. To illustrate this issue, consider a top-down analysis of the dynamics of the theory. We thus have a classical space-time as a block of atoms of space, and all these atoms can in turn be grouped into distinct configurations piled together according to (\ref{6}). BLQG then selects a privileged form of carrying out this task. The reason is the ontological commitment of BLQG to exactly one actual initial configuration of atoms of space and exactly one path from an initial to a final configuration that is selected by the guiding equation (\ref{6}), which is a non-local law, always taking one entire configuration to a subsequent entire configuration. Consequently, there is a fact of the matter which of these entire configurations are connected by the law (\ref{6}) in any world of BLQG.\\
Thus, it may turn out that the ontological clarity of the Bohmian approach to any quantum theory comes at the price of being committed to more geometrical structure of space-time than is admitted in relativity physics - with, however, it being possible that this additional structure is provided for by the universal wave-function (for a suggestion in that sense, see \citealp{bohrel}). In other words, it may be the case that the Bohmian approach is committed to maintaining that when considered from a quantum perspective, one of the main tenets of relativistic theories, viz. the absence of a privileged space-time foliation, proves to be an epistemic instead of an ontic affair. This means that we cannot have knowledge of the additional quantum-geometrical structure involved because we cannot know the initial configuration of the elements of the primitive ontology (the initial configuration of atoms of space in BLQG). However, endorsing a privileged quantum foliation of space-time is compatible with all the empirical results of GR. If one rejects that commitment, then one has to do better and to show how one can formulate a clear ontology for a quantum theory of matter and space-time that pays heed to quantum non-locality without subscribing to a privileged foliation of space-time. In any case, it is not an admissible attitude to refuse to enter into the business of setting out a clear ontology for quantum physics because doing so may bring out a tension between quantum physics and relativity physics.\\
Our proposal for a Bohmian theory of LQG provides a precise sense in which space-time is emergent and in which it is not emergent (see \citealp{314}, for a philosophical discussion of the claim of the emergence of space-time in QG): space-time is emergent in that it is built up from a series of configurations of atoms of space and two laws, namely, (\ref{5}) that selects all the actualizable configurations in a possible world and (\ref{6}) that provides for a dynamics of the transition from one such configuration to subsequent ones. But space-time emerges from concrete physical entities, namely atoms of space whose initial configuration approximates a classical $3$-dimensional space. It is thus much more precise to talk in terms of the approximation of the space-time of GR by configurations of elements of the primitive ontology of BLQG ordered by a certain law than to use the expression ``emergence of space-time''. There is no problem in BLQG how a quantum configuration can lead to a classical space-time, since according to the Bohmian approach, the objects of the theory are in any case classical in that there always is exactly one actual configuration of the elements of the primitive ontology (instead of superpositions) and all what is specific for a quantum theory is contained in the law that applies to these configurations. Due to its ontological precision, the Bohmian approach to QG does not contain any unclear - and unintelligible - claims that confuse means of mathematical representation with the ontology of a physical theory in suggesting that space-time could emerge from non-spatial entities such as a wave-function in configuration space, algebraic relations among operators, etc.\\
In conclusion, whatever the truth of the matter may be, as in the non-relativistic case, so also in the case of QG, the Bohmian approach sets a standard that any proposal for a serious ontology in this field should meet, distinguishing between the ontology of a physical theory and the means of its mathematical representation as well as providing a link from the fundamental ontology to classical phenomena (such as the classical space-time of GR).

\pdfbookmark[1]{Acknowledgements}{acknowledgements}
\begin{center}
\textbf{Acknowledgements}:
\end{center}
We wish to thank an anonymous referee and Christian W\"uthrich for useful comments on an earlier version of the paper. One of us (A.V.) gratefully acknowledges partial financial support from the ``Fondation du 450e pour l'UNIL''.

\pdfbookmark[1]{References}{references}
\bibliography{Ref}
\end{document}